\documentclass[journal]{IEEEtran}
\usepackage{color}
\usepackage{cite}
\usepackage{graphicx}
\usepackage{multirow}
\usepackage{epstopdf}
\usepackage{booktabs}
\usepackage{amsmath}
\usepackage[T1]{fontenc}
\usepackage{authblk}
\usepackage{verbatim}
\usepackage{breqn}
\begin{document}
%
\title{Accurate Localization in Wireless Sensor Networks in the Presence of Cross Technology Interference}

\author{Usman Nazir\\
Department of Computer Science, Syed Babar Ali School of Science and Engineering\\Lahore University of Management Sciences (LUMS), Lahore, Pakistan \\
{\tt\small 17030059@lums.edu.pk}}

\maketitle
\pagenumbering{gobble}


\begin{abstract}
Localization of mobile nodes in a wireless sensor networks (WSNs) is an active area of research. In this paper, we present a novel RSSI based localization algorithm for 802.15.4 (ZigBee) based WSNs. We propose and implement a novel range based localization algorithm to minimize cross technology interference operating in the same band. The goal is to minimize the mean square error of the localization algorithm. Hardware implementation of the algorithm is in agreement with ideal (no interference) simulation results where an accuracy of less than 0.5m  has been achieved.


\end{abstract}


\IEEEpeerreviewmaketitle

\section{Introduction}
\begin{table*}[t]
 \centering
  \caption{Specifications of SYNAPSE RF266 Module}
    \begin{tabular}{|p{3.5cm}|p{4cm}|}
    \hline

    Parameters & Value \\ \hline
    Indoor Range& Up to 200ft. at 250kbps\\
    Outdoor LOS Range&	Up to 4000ft. at 250kbps\\
    Transmit Power Output&	20dBm\\
    RF Data Rate&	250kbps, 500kbps, 1Mbps,2Mbps\\
    Receiver Sensitivity&	-107 dBm\\
    Supply Voltage&	2.7 - 3.6 V \\
    Transmit Current (Typ)&	130 mA\\
    Receive Current (Typ)&	25 mA\\
    Sleep Current (Typ)&	1.18 $\mu$A (internel timer off) \& 2.3 $\mu$A (internel timer on)\\
    Topology&	Mesh (SNAP)\\
    Number of Channels&	16\\
    \hline
\end{tabular}%
  \label{tab:spec}%
\end{table*}%
\IEEEPARstart{P}{}hysical location of mobile nodes is often required for a large number of applications in wireless sensor networks (WSNs). Localization techniques can be broadly classified into range free or range based, anchor free or anchor based and distributed  or centralized  techniques.  Centralized localization techniques \cite{C.Luo, Y.Sha, Y.Shar, A.Kan, C.Ali, R.C} transfer the entire data to a centralized node, where the localization algorithm estimates the position for all of the mobile nodes. These techniques have a lot of communication overhead. In the distributed localization techniques
\cite{C.Luo, D.Nic, M.Jin, K.Lan, S.Nav, T.He, A.Sav, S.Sim, J.Bac, B.Pri, D.Moo, L.Mee, K.Yip, A.Ahm, M.Mar, N.Pat, R.Hua, A.Als, J.Llo, R.C}, the mobile nodes are themselves capable of estimating their location. The location of the mobile node is transferred to the central node only in case of an event or with the sink initiated query, depending on the network design, thereby reducing the communication cost significantly. Anchor free techniques \cite{M.Jin, Y.Shar, J.Bac, B.Pri, D.Moo, L.Mee, A.Ahm, N.Pat} do not requires beacon signals from the anchor nodes but offer very limited localization accuracy whereas anchor based techniques
\cite{C.Luo, D.Nic, K.Lan, S.Nav, Y.Sha, A.Kan, C.Ali, T.He, A.Sav, S.Sim, K.Yip, A.Ahm, M.Mar, R.Hua, A.Als, J.Llo, R.C} require beacon signals from anchor nodes of known location. Range free techniques
\cite{M.Jin, K.Lan, S.Nav, Y.Sha, Y.Shar, A.Kan, T.He, S.Sim, J.Bac, B.Pri, D.Moo, L.Mee, K.Yip} rely on attributes such as hop count, connectivity information etc. Although, these are cost effective techniques but localization results are not accurate.  The Range based techniques
\cite{C.Luo, D.Nic, K.Lan, C.Ali, A.Sav, A.Ahm, M.Mar, N.Pat, R.Hua , A.Als, J.Llo, R.C} rely on received signal strength indicator (RSSI), time of arrival (ToA), angle of arrival (AoA), time difference of arrival (TDoA) etc. ToA, AoA and TDoA, all have better localization accuracy as compared to RSSI based techniques but require additional hardware for their implementation.

The RSSI based localization schemes are by far the most popular localization techniques employed for monitoring location of mobile sensor nodes as it requires minimal, low cost hardware for its implementation.  In RSSI based localization implementations, RF transceivers capable of measuring RSSI are used for location estimation. Multiple technologies for instance IEEE 802.15.4 (ZigBee), IEEE 802.15.1 (Bluetooth) and IEEE 802.11 (WiFi) all share the same 2.4 GHz ISM band~\cite{C.Nod}. The interference caused by these technologies causes RSSI to fluctuate rapidly, causing packet collisions and affecting the localization accuracy. Addressing cross technology interference (CTI) is therefore critical for systems operating in license free ISM bands. The highest CTI for a Zigbee based radio is caused by Wifi signals, because Wifi is often co-located with IEEE 802.15.4 networks.

\begin{figure}[!t]
\begin{center}
\includegraphics[scale=0.20]{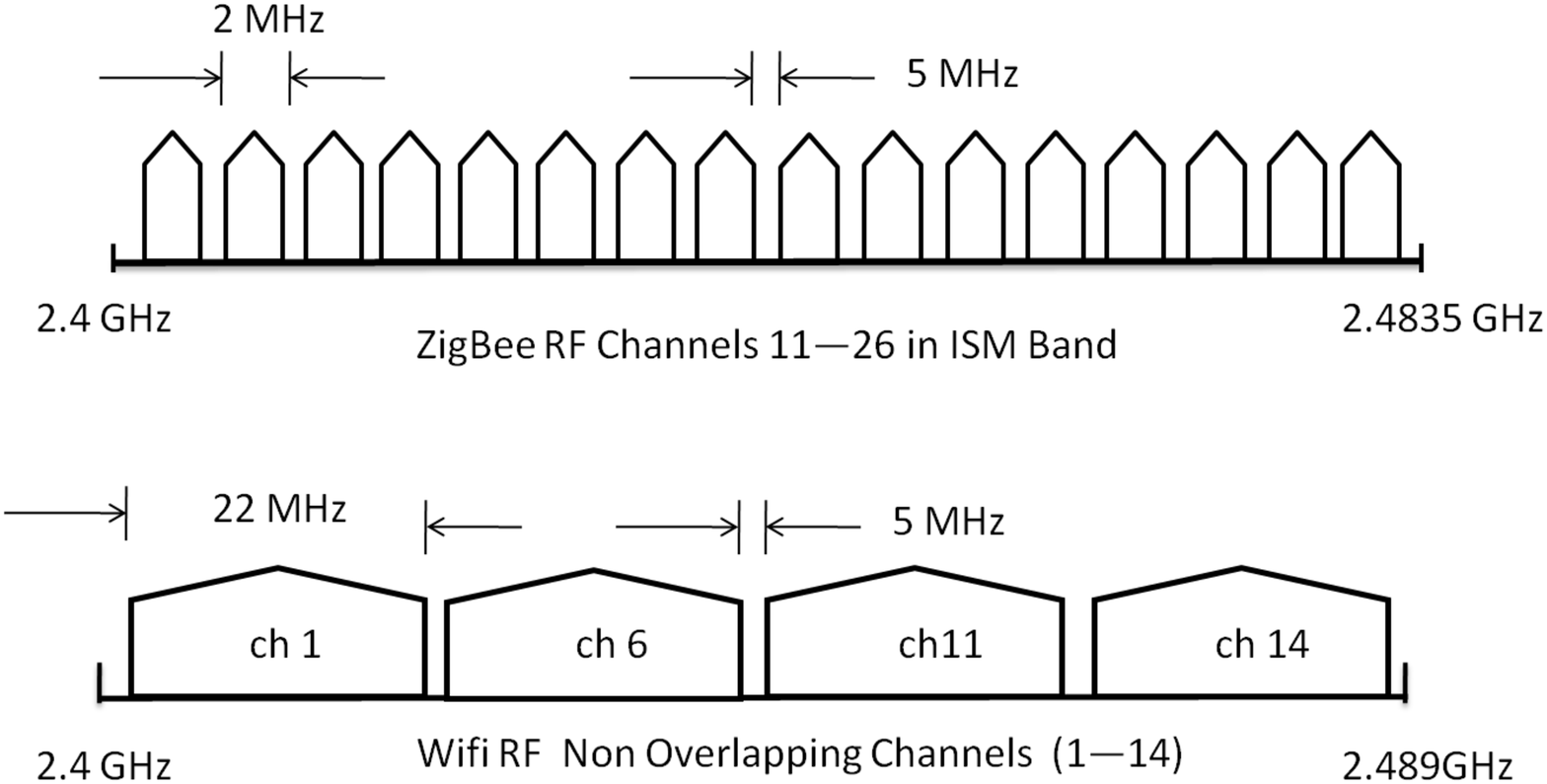}
\end{center}
\vspace{-0.4cm}
\caption{ZigBee and WiFi RF channels in 2.4 GHz band. ZigBee channel is 2 MHz wide and WiFi channel is 22 MHz wide}
\vspace{0.2cm}
\label{fig:zc}
\end{figure}

In order to minimize RF interference at the receiver front end, we use aggregation function to calculate average RSSI values. We have used Synapse RF266 sensor nodes having 802.15.4 radios and specifications shown in Table~\ref{tab:spec}. Detailed specifications can be seen in ~\cite{web}. ZigBee and WiFi radio channels are shown in Fig.~\ref{fig:zc}. As shown, the bandwidth of each ZigBee and Wifi channel is 2MHz and 22MHz respectively, whereas separation between two adjacent ZigBee and Wifi Channels is 5MHz~\cite{N.C.Tas}. In Wifi band, non overlapping channels are used in Europe and North America~\cite{G.Yan}. Typically, channels with high interference have higher mean RSSI values~\cite{R.Mus}. Therefore, ZigBee channel with the lowest average RSSI value is selected to acquire minimal RF interference  with Wifi signals \cite{R.Mus}. In addition to RF interference, the received signal also gets affected by White Gaussian noise at the receiver front end which has an infinite spectrum and thus infinite energy.  To cater this problem, we introduce Kalman filtering at the receiver front end. We show that if RSSI after being processed for interference minimization and noise filtering is given as an input to a trilateration localization algorithm, mean square error of the localization algorithm is minimized.

The main contributions of this paper have been summarized as follows:

\begin{itemize}
	\item We introduce a novel cross technology interference minimization scheme with trilateration for an improved accuracy.
	\item We propose the use of Kalman filter at the receiver front for noise minimization of localization algorithms.
	\item Hardware implementation of the localization algorithm with minimal CTI and noise reduction.
\end{itemize}

The remainder of the paper is organized as follows. Section II explains RSSI based localization algorithm along with CTI minimization and noise reduction techniques. Complete system design with hardware implementation has been presented in Section III. Discussion on experimental results has been elaborated in Section IV. Node deployment strategy is discussed in section V and Section VI outlines the future directions and concludes this paper.

\section{Proposed Localization Algorithm with Kalman Filtering}

As discussed in Section I, localization algorithms can be classified in the following ways:
\begin{enumerate}
  \item Centralized or Distributed Algorithms
  \item Range free or Range based Algorithms
  \item Anchor free or Anchor based Algorithms
\end{enumerate}

Each type of localization technique has its advantages or disadvantages. A detailed discussion on the merits and demerits of localization techniques is out of scope of this paper, however a summary of the state of the art for localization algorithms has been provided in Table~\ref{tab:num1}. In this paper we make use of a distributed, range and anchor based localization algorithm to localize a target (mobile) node.
\subsection{Log Distance Path Loss Model}
\begin{figure}[h]
\begin{center}
\includegraphics[scale=0.20]{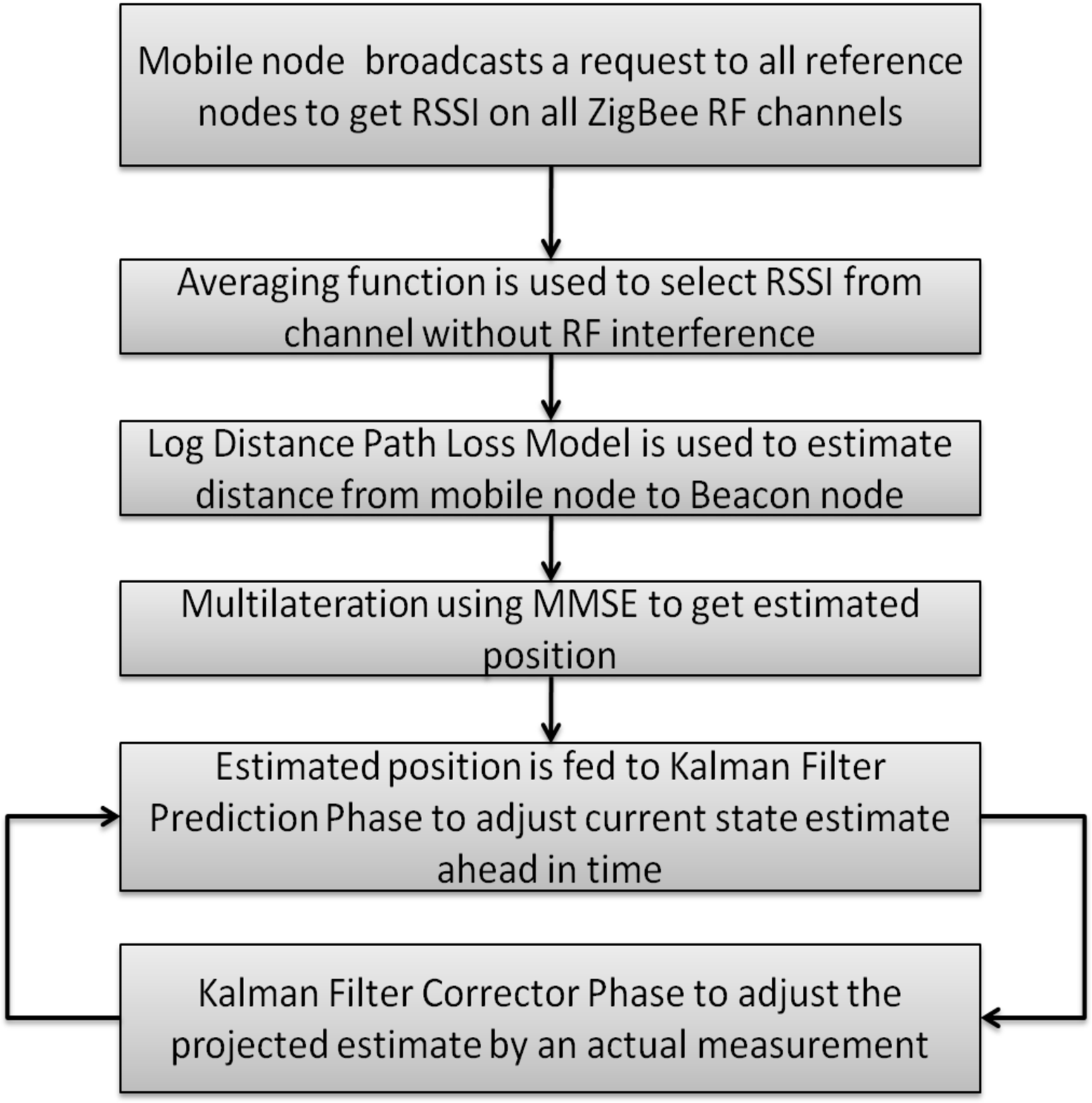}
\end{center}
\vspace{-0.4cm}
 \caption{Proposed Localization Algorithm}
\vspace{0.2cm}
\label{fig:sd}
\end{figure}

The algorithm runs on the mobile node where RSSI of neighboring nodes is needed. In our algorithm we use aggregation function, which averages the received RSSI over multiple acquisitions, to verify the the RF interference between ZigBee and Wifi Channels. The ZigBee channel with lowest mean RSSI Value gets selected for communication. RSSI values of neighboring nodes are acquired at the mobile node over the selected frequency (channel). The mobile node then makes use of the log distance path loss model, given below in equation~\ref{eq:rssi}, to calculate its distance to the neighboring transmitting nodes.

 \begin{equation}\label{eq:rssi}
    RSSI{(d)} =  RSSI{(d_{o})}-10n \times log{(d/d_{o})}
\end{equation}

where $d$ is the distance between transmitter and the receiver, $d_{0}$ is a reference distance which we is assumed to be 1 meter. $RSSI{(d_{0})}$ is the RSSI at reference distance taken as -45dBm and $RSSI{(d)}$ is the received signal strength at distance $d$.
Since a Trilateration approach~\cite{Y.T.Che} has been used to estimate the location of sensor node, RSSI value for at least three neighboring nodes is required for a valid location estimate of the mobile node. Once the location estimate from trilateration algorithm is received, a Kalman Filter is used to minimize the mean squared error. A flow chart of the the localization process has been presented in the Fig.~\ref{fig:sd}.

\subsection{Trilateration}

In a trilateration algorithm along with the minimum mean square estimation (MMSE) ~\cite{Y.T.Che}~\cite{X.Li}, the position of the mobile node can be computed by solving the following equation:

  \begin{equation}\label{eq:trilateration}
\hat{P}=\left(\!
    \begin{array}{c}
      \hat{x} \\
      \hat{y}
    \end{array}
  \!\right) = (A^TA)^{-1}A^TB
\end{equation}

where $\hat{P} = (\hat{x},\hat{y})^T$ denotes the estimated target location, and the matrices A and B are defined as
\begin{equation}
A = \left(\begin{IEEEeqnarraybox*}[][c]{,c/c,}
2(x_{2}-x_{1})&2(y_{2}-y_{1})\\
2(x_{3}-x_{1})&2(y_{3}-y_{1})
\end{IEEEeqnarraybox*}\right)
\end{equation}
\begin{equation}
B = \left(\begin{IEEEeqnarraybox*}[][c]{,c/c,}
d_{1,target}^2-d_{2,target}+(x_{2}^2+y_{2}^2-x_{1}^2-y_{1}^2)\\
d_{1,target}^2-d_{3,target}+(x_{3}^2+y_{3}^2-x_{1}^2-y_{1}^2)
\end{IEEEeqnarraybox*}\right)
\end{equation}

where $(x_{i},y_{i})$ gives the position of the beacon nodes. Note that for the implementation of trilateration algorithm, position of at least three beacon nodes is required.
\subsection{Kalman Filtering}

The estimated position, $\hat{P}$, is then filtered through a Kalman filter to reduce the effects of the noise. The Kalman filter implementation has two phases. The first phase is the prediction phase where the initial current state estimate and the error covariance are computed. The  initial current state can be calculated by using the estimated location by the following equation:

\begin{equation}\label{eq:Kal1}
\hat{\hat{P}} = S_{T}\hat{P} + \Pi
\end{equation}

where $\hat{\hat{P}} = (\hat{\hat{x}},\hat{\hat{y}})^T$ is the initial current state estimate computed from the previous estimate $\hat{P}$, $S_{T}$, $S_T$ is state transition matrix and $\Pi$ is the control matrix initialized to zero. The error covariance matrix can then be computed as

\begin{equation}\label{eq:Kal2}
\hat{E} =S_{T}E_{int}S_{T}^{T}+Q
\end{equation}

where $E_{int}$ is the initial error covariance (supposed to be zero in our case), and $Q$ is process noise covariance matrix.


The second phase of the Kalman filter implementation is the corrector phase and is used to estimate actual position ($P$).  Kalman corrector phase is mathematically given by the following equation:

\begin{equation}\label{eq:Kal3}
P=\hat{\hat{P}} +K_k (z_{k}-H\hat{\hat{P}}) \\
\end{equation}
where $K_k$ is Kalman gain expressed in eq. (\ref{eq:kalgain}), $H$ is the observation matrix and $z_{k}$ is the measurement vector given by following equation

 \begin{equation}\label{eq:MV}
z_k = \left(\!
    \begin{array}{c}
     \sqrt{(\hat{x}-x_{1})^{2} + (\hat{y}-y_{1})^{2}}\\
     \sqrt{(\hat{x}-x_{2})^{2} + (\hat{y}-y_{2})^{2}}\\
    \sqrt{(\hat{x}-x_{3})^{2} + (\hat{y}-y_{3})^{2}}
    \end{array}
  \!\right)
\end{equation}

 \begin{equation}\label{eq:kalgain}
 K_{k}= \frac{\hat{E}H^{T}}{H\hat{E}H^{T}+R}\\
\end{equation}

In equation~\ref{eq:kalgain}, $R$ is the estimated measurement error covariance (environmental noise) and $\hat{E}$ is the predicted error covariance. The environmental noise matrix describes the noise inferred on data from the sources that lie within the path from the sensed object to the filter. Error covariance is iteratively updated using the equation~\ref{eq:EC}.

 \begin{equation}\label{eq:EC}
 \hat{E}=(1-K_{k}H)\hat{E}
\end{equation}

\begin{table*}[t]
 \centering
  \caption{Classifications of Proposals for Localization in WSNs \cite{U.Naz}}
    \begin{tabular}{|p{5cm}|p{3.5cm}|p{3.5cm}|p{3cm}|}
    \hline

    PROPOSALS & RangeBased$/$RangeFree & AnchorBased$/$AnchorFree & Distributed$/$Centralized \\ \hline
    Ren C. Luo Fellow, et al.\cite{C.Luo}& RangeBased & AnchorBased & Both                        \\ \hline
    D. Niculesu and B. Nath\cite{D.Nic}& RangeBased & AnchorBased & Distributed                                  \\ \hline
    Miao Jin, et al.\cite{M.Jin}& RangeFree & AnchorFree & Distributed                                 \\ \hline
    Koen Langendoen, et al.\cite{K.Lan}& Both & AnchorBased & Distributed                                \\ \hline
    E. S. Navarro, et al. \cite{S.Nav}& RangeFree & AnchorBased & Distributed \\ \hline
    Yi Shang, et al.\cite{Y.Sha}& RangeFree & AnchorBased & Centralized \\ \hline
    Yi Shang, et al.\cite{Y.Shar}& RangeFree & AnchorFree & Centralized \\ \hline
    Anushiya A Kannan, et al.\cite{A.Kan}& RangeFree&AnchorBased&Centralized\\ \hline
    C. Alippi, et al.,\cite{C.Ali}&RangeBased&AnchorBased&Centralized\\ \hline
    T. He, et al.,\cite{T.He}&RangeFree&AnchorBased&Distributed \\ \hline
    A. Savvides, et al.\cite{A.Sav}&RangeBased&AnchorBased&Distributed \\ \hline
    S. Simic, et al.\cite{S.Sim}&RangeFree&AnchorBased&Distributed \\ \hline
    J. Bachrach, et al.\cite{J.Bac}&RangeFree&AnchorFree&Distributed \\ \hline
    Nissanka B. Priyantha, et al.\cite{B.Pri}&RangeFree&AnchorFree&Distributed \\ \hline
    D. Moore, et al.,\cite{D.Moo}& RangeFree&AnchorFree&Distributed \\ \hline
    L. Meertens, et al.,\cite{L.Mee}&RangeFree&AnchorFree&Distributed \\ \hline
    King-Yip Cheng, et al.,\cite{K.Yip}&RangeFree&AnchorBased&Distributed \\ \hline
    A. A. Ahmed, et al.,\cite{A.Ahm}, &RangeBased&Both&Distributed \\ \hline
    M. Maroti, et al.,\cite{M.Mar}, &RangeBased&AnchorBased&Distributed \\ \hline
    Neal Patwari, et al.\cite{N.Pat},&RangeBased&AnchorFree&Distributed \\ \hline
    R. Huang, et al.,\cite{R.Hua}, &RangeBased&AnchorBased&Distributed \\ \hline
    N. A. Alsindi, et al.,\cite{A.Als}&RangeBased&AnchorBased&Distributed \\ \hline
    J. Lloret, et al.,\cite{J.Llo}&RangeBased&AnchorBased&Distributed \\ \hline
    Ren C. Luo Fellow, et al.\cite{R.C}&RangeBased&AnchorBased&Both \\ \hline

\end{tabular}%
  \label{tab:num1}%
\end{table*}%

\section{Hardware Implementation of the Proposed Algorithm}
In this section, we describe the complete software and hardware implementation of localization algorithm. We make use of the wireless programming feature of the off the shelf SYNAPE sensor mote to program them Over-the-Air. The software used for programming is named as `portal'  and is a the SYNAPSE designated software. All the nodes can be programmed simultaneously via portal. Each node is capable of receiving RSSI values from its neighboring nodes, distributed in the region of interest (ROI).

Remote Procedure Calls (RPCs) are transmitted by mobile (target) node on all of ZigBee channels. 
 The mobile node then broadcasts a request to all reference nodes which are in range of the target node to receive their RSSI values. Thereafter, the target node tries to find out the channel with minimum RF interference from Wifi radio transmitters co-located in the environment. The built in functions of SYNAPSE motes can be used to find out the  channel from which the target is receiving minimum mean energy. The RSSI value for each anchor or beacon node is measured at $100ms$ intervals and then averaged out over a delay of $1sec$. The aggregated RSSI value is then fed into the localization algorithm.


\subsection{Channel Selection}
\begin{figure}[h]
\begin{center}
\includegraphics[scale=0.20]{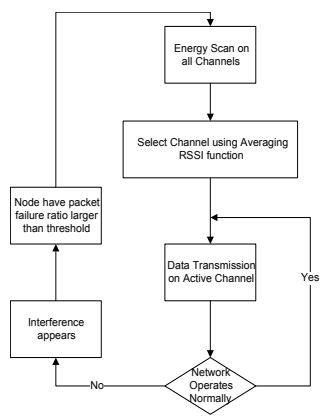}
\end{center}
\vspace{-0.4cm}
 \caption{Procedure of Channel Selection}
\vspace{0.2cm}
\label{fig:scanning}
\end{figure}

In order to minimize CTI, channels receiving minimum RF energy gets selected. The procedure for channel selection has been shown in Fig.~\ref{fig:scanning}.  First of all, the target node scans energy on all the ZigBee channels and acquires RSSI with some sampling rate. The acquired values are then averaged to minimize the effect of random fluctuations. Thereafter, the target node selects the channel with minimum interference and initializes this channel in active mode to establish communication with anchor nodes. Target node monitors this active channel continuously for any possible interference. As soon as the interference is detected on this active channel, i.e. the packet failure ratio crosses a certain set threshold, the process of channel selection gets repeated (see Fig. \ref{fig:scanning}).


\subsection{Implementation of Localization Algorithm}
\begin{figure}[h]
\begin{center}
\includegraphics[scale=0.20]{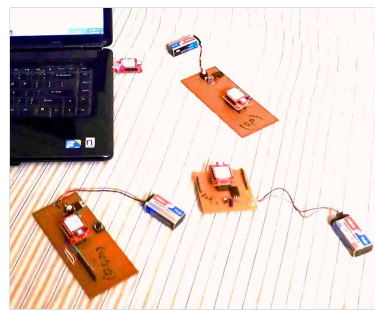}
\end{center}
\vspace{-0.4cm}
 \caption{Experimental Setup}
\vspace{0.2cm}
\label{fig:hw}
\end{figure}

As mentioned previously, the localization algorithm has been implemented using off the shelf SYNAPSE sensor motes \cite{web}. Experimental setup has been presented in the Fig. \ref{fig:hw} where the target node is attached to a Laptop where SYNAPSE Portal software is running. Target node scans all ZigBee channels using built in functions and selects the channel with minimum interference, as described previously.  After selection of channel with minimum RF interference, target nodes communicates to its neighbors and acquires the RSSI values of at least three neighbors (beacon) nodes. Thereafter, trilateration localization technique is used to calculate the position of target node. \\
The mobile node can also be programmed without connecting it to the laptop using SNAPpy language through the Portal's Over-the-Air programming feature. The localization algorithm however, remains the same. The target first automatically scans the ZigBee channels and selects one channel for communication with beacon nodes. After getting average RSSI values from at least three beacons it implements trilateration algorithm to calculate its position and uses Kalman Filter for improved position accuracy.

\section{Experimental Results and Discussions}

\begin{figure}[h]

\begin{center}
\includegraphics[scale=0.20]{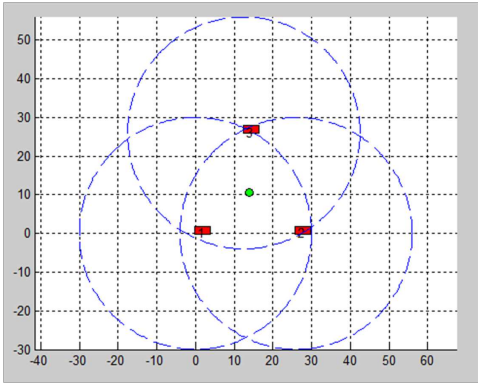}
\end{center}

\vspace{-0.4cm}
 \centering\caption{Localization using Trilateration Approach}
\vspace{0.2cm}
\label{fig:result}
\end{figure}

A snapshot of trilateration localization algorithm, implemented in the absence of Wifi interference with the SYNAPSE sensor motes, has been shown in Fig.~\ref{fig:result}, where green is the target node and red nodes are anchor nodes.  If the target node receives RSSI from more than three neighbor nodes, target node selects three neighbors with maximum RSSI. Log Distance Path Loss model (eq.~\ref{eq:rssi}) is then used to find out its  distances to anchor nodes and its location using trilateration algorithm. Thereafter, Kalman filter is implemented using equations~\ref{eq:Kal1} and~\ref{eq:Kal2} to minimize the mean  squared error of the estimated position and improve the localization accuracy. As shown in Fig.~\ref{fig:result}, anchor nodes have positions (0,0), (30,0) and (15, 30) in cartesian coordinates respectively. Target node receives RSSI of $-60 $ dBm from all anchor nodes where as the reference RSSI is measured to be $-45 $ dBm at $1$m distance. 

\begin{figure}[h]
\begin{center}
\includegraphics[scale=0.60]{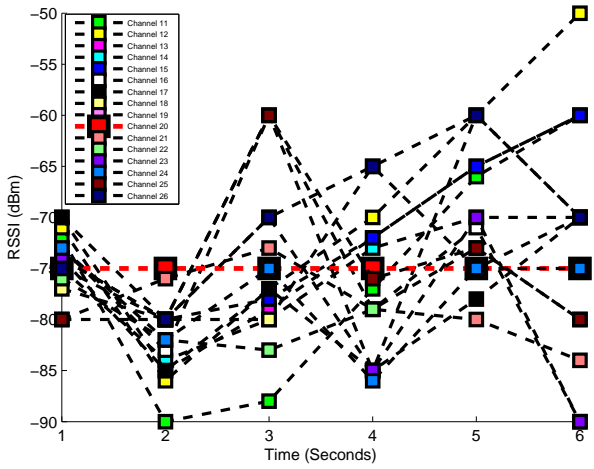}
\end{center}
\vspace{-0.4cm}
 \caption{RSSI Variations on ZigBee Channels}
\vspace{0.2cm}
\label{fig:result2}
\end{figure}

The experiments are then conducted in the presence of RF interference from Wifi transceivers.  Using the built in channel analyzer tool for ZigBee motes, RSSI values, prior to any transmission, detected on all channels have been shown in Fig.~\ref{fig:result2}. RSSI values for channels 11-26 have been shown in the figure but channel 20 has lowest variations in RSSI values w.r.t. time. Therefore, channel 20 has been selected for Zigbee transmission by the target node to communicate with the anchor nodes. Once the channel gets selected, trilateration localization algorithm is invoked in a similar manner as in the case of no interference.

\begin{figure}[h]
\begin{center}
\includegraphics[scale=0.60]{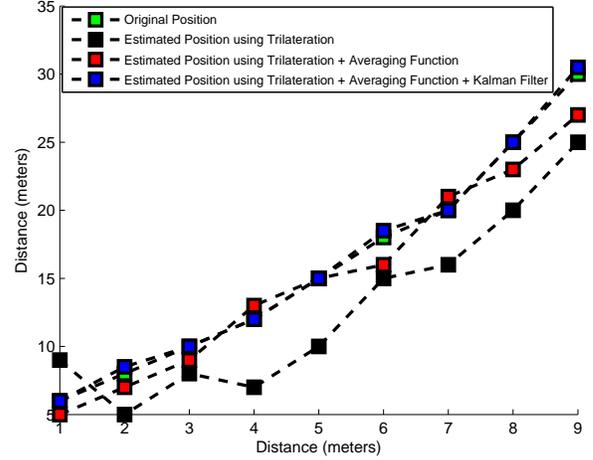}
\end{center}
\vspace{-0.4cm}
 \caption{Original Position and Estimated Position of Mobile Node}
\vspace{0.2cm}
\label{fig:result1}
\end{figure}

Fig.~\ref{fig:result1} presents the results of the trilateration algorithm with and without applying the averaging function. The curve also presents the result of localization algorithm after employing both averaging as well as Kalman filter. The deviation of the estimated position from the original position can be compared for different schemes. If we simply use Trilateration technique to estimate target position, accurate estimation of the target position is not possible. If we find estimated position of target node using Trilateration technique with the averaging function, we get a better estimate of the original position but still the deviation cannot be ignored (see Fig.~\ref{fig:result1}). Finally,  if we apply Kalman filter along with the averaging function, the estimated and the original position are in good agreement with each other. Thus, with Kalman filtering, we get minimal deviation between estimated position and original positions. The result show that we achieve a localization accuracy of less than  0.5m with proposed algorithm.


%
%
\section{Node Deployment Strategy}
\begin{figure}[!t]
\begin{center}
\includegraphics[scale=0.60]{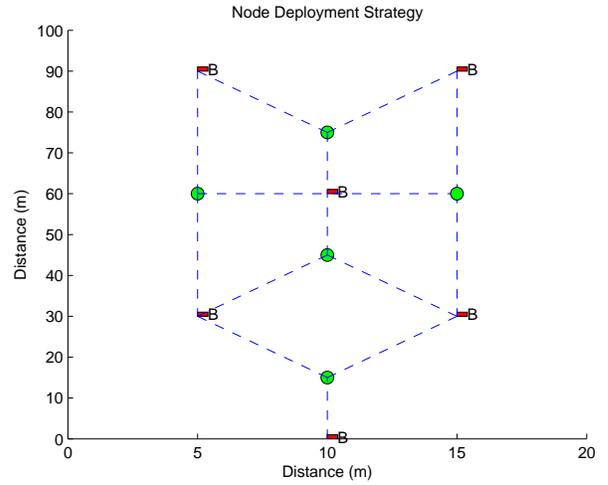}
\end{center}
\vspace{-0.4cm}
\caption{Node Deployment Strategy}
\vspace{0.2cm}
\label{fig:deploy}
\end{figure}

Based on the experimental results, we proposed a nodes deployment strategy for indoor environment. With this deployment strategy, mobile node will always have connectivity with at least three beacon nodes to fulfill the requirement of trilateration based localization technique. As shown in Fig. \ref{fig:deploy}, where green is the target node and red nodes are beacon nodes. If the target node receives RSSI from more than three neighbor nodes, target node selects three neighbors with maximum RSSI. Log Distance Path Loss model (eq.~\ref{eq:rssi}) is then used to find out its distances to anchor nodes and its location using trilateration algorithm. Using this deployment strategy in the region of interest, mobile node will always be in the range of at least three beacon nodes.

\section{Conclusion and Future Work}
In this paper, a novel RSSI based localization technique has been introduced and implemented. The proposed scheme minimizes cross technology interference (CTI)and apply noise reduction filters to achieve very high localization accuracy. Hardware implementation of the proposed scheme has been carried out and the result suggest that scheme can easily be implemented on Zigbee based sensor  motes. The combined CTI minimization and noise reduction technique achieves a  localization accuracy of less than 0.5m on commercial off the shelf sensor motes. Future work includes addressing CTI in networks using ISM band with the help of Machine Learning algorithms and Kalman Filtering. Accurate localization implementation over DASH7 sensor motes is also a work in progress.


\bibliographystyle{IEEEtran}
\bibliography{IEEEabrv,main}


\end{document}